# Proton tracking in a high-granularity Digital Tracking Calorimeter for proton CT purposes

H. E. S. Pettersen*[a, b], J. Alme[b], A. Biegun[e], A. van den Brink[c], M. Chaar[b], D. Fehlker[b], I. Meric[d], O. H. Odland[a], T. Peitzmann[c], E. Rocco[c], K. Ullaland[b], H. Wang[c], S. Yang[b], C. Zhang[c], D. Röhrich[b]

*Corresponding author: E-mail: helge.pettersen@helse-bergen.no.
[a]Department of oncology and medical physics, Haukeland University Hospital, Postbox 1400, 5021 Bergen, Norway
[b]Department of physics and technology, University of Bergen, Postbox 7803, 5020 Bergen, Norway
[c]Nikhef, Utrecht University, Postbox 41882, 1009 DB Amsterdam, the Netherlands
[d]Department of Electrical Engineering, Bergen University College, Postbox 7030, 5020 Bergen, Norway
[e]Kernfysisch Versneller Instituut, University of Groningen, NL-9747 AA Groningen, the Netherlands

## Abstract

Radiation therapy with protons as of today utilizes information from x-ray CT in order to estimate the proton stopping power of the traversed tissue in a patient. The conversion from x-ray attenuation to proton stopping power in tissue introduces range uncertainties of the order of 2-3% of the range, uncertainties that are contributing to an increase of the necessary planning margins added to the target volume in a patient. Imaging methods and modalities, such as Dual Energy CT and proton CT, have come into consideration in the pursuit of obtaining an as good as possible estimate of the proton stopping power. In this study, a Digital Tracking Calorimeter is benchmarked for proof-of-concept for proton CT purposes. The Digital Tracking Calorimeter was originally designed for the reconstruction of high-energy electromagnetic showers for the ALICE-FoCal project. The presented prototype forms the basis for a proton CT system using a single technology for tracking and calorimetry. This advantage simplifies the setup and reduces the cost of a proton CT system assembly, and it is a unique feature of the Digital Tracking Calorimeter concept. Data from the AGORFIRM beamline at KVI-CART in Groningen in the Netherlands and Monte Carlo simulation results are used to in order to develop a tracking algorithm for the estimation of the residual ranges of a high number of concurrent proton tracks. High energy protons traversing the detector leave a track through the sensor layers. These tracks are spread out though charge diffusion processes. A charge diffusion model is applied for acquisition of estimates of the deposited energy of the protons in each sensor layer by using the size of the charge diffused area. A model fit of the Bragg Curve is applied to each reconstructed track and through this, estimating the residual range of each proton. The range of the individual protons can at present be estimated with a resolution of 4%. The readout system for this prototype is able to handle an effective proton frequency of 1 MHz by using 500 concurrent proton tracks in each readout frame, which is at the high end range of present similar prototypes. A future further optimized prototype will enable a high-speed and more accurate determination of the ranges of individual protons in a therapeutic beam.

**Keywords**[1]
**Abbreviations**[2]

---

[1]Proton Therapy, Proton CT, Monte Carlo, Particle Tracking, Calorimeter, Proton Range Estimation
[2] DAQ: Data Acquisition; DTC: Digital Tracking Calorimeter; ENC: Equivalent Noise Charge; GATE: Simulation Software for Geant4; HU: Hounsfield Unit; MC: Monte Carlo; ROOT: Data Analysis Framework; RSP: Relative Stopping Power; WET: Water Equivalent Thickness.

# 1 Introduction

There has been a significant increase in the number of cancer patients treated with proton radiation therapy in the recent decades worldwide. As of January 2015, more than 137 000 patients have been treated with charged particle therapy [1]. The motivations for application of proton therapy during cancer treatment are the prospects of reducing the irradiated volume of the patient during radiation treatment. Short term and long term treatment-induced side effects, such as the probability for radiation-induced secondary cancers, are then reduced due to the finite range of protons in tissue.

Proton therapy as of today is performed with the delivery of pre-calculated dose plans for each patient: The applied dose plans are based on x-ray computed tomography (CT) images. The x-ray CT images reflect the patient's anatomy, however they provide limited resolution for calculating how the protons traverse and deposit dose in the patient's body during proton therapy. The Relative Stopping Power (RSP) for protons in tissue is needed in order to calculate the proton range during dose calculations in a Treatment Planning System. The RSP is obtained by converting attenuation of x-rays, represented by Hounsfield Unit (HU) maps in tissue, to RSP maps in the same tissue, through pre-determined HU-to-RSP conversion curves [2]. This conversion procedure introduces range uncertainties in the order of 2-3% [3], corresponding to 4-6 mm at treatment depth 20 cm into the patient. The dual goal during radiation therapy can be expressed as to irradiate the tumor with the described treatment dose at the same time as one is limiting the amount of healthy tissue irradiated to a minimum. Thus, keeping the margins as narrow as possible is an important goal during radiation treatment. The uncertainties introduced by the HU-to-RSP conversion, as well as by other treatment-specific and particle-specific uncertainties, such as the impact on the range in tissue due to unavoidable variations in the density composition along each proton's path in tissue, necessitates the use of *robust* proton treatment planning [4]. Robust treatment planning takes a set of uncertainties into account when optimizing the size and shape of the Planned Target Volume to be irradiated, which in turn has a consequence that during proton therapy one normally does not fully utilize the possibilities of the sharp distal dose gradient inherent in the physical properties of the proton's interaction with matter.

A proton CT system would yield a direct link to calculation of the RSP map in the patient, thus avoiding the indirect deduction of the RSP values based upon the HU-to-RSP conversion factors. Proton therapy can be further applied in an optimized and enhanced fashion by a more precise knowledge of the range of protons in matter. With a proton CT system, the dose planners will be able to apply margins that are at the same time clinically safe, but also limited downwards to the best of knowledge and technology level, thus avoiding unnecessary irradiation of healthy tissue.

During a proton CT scan, a high-energy proton beam is directed at the patient: the proton beam must have sufficient energy to completely pass through the object (the patient) being imaged. The proton residual energy is measured after the protons have traversed through the patient and into a detector placed distal to the patient, as seen from the proton beam delivery system. The residual energy from each proton can then used, together with the proton's estimated path through the patient, as a basis for reconstructing a volume with RSP. The RSP can be used as an input for patient dose planning software. Other kinds of information output from this imaging technique are also feasible: attenuation maps applied for measuring the nuclear interaction cross sections [5], multiple scattering effects [6] and range straggling distributions.

Telescopic ionization chambers or calorimeters measure, respectively, the remaining range or energy of each individual proton after traversing the patient. Tracking detectors proximal and distal to the patient yield information needed in order to obtain a measure of the path of each proton through the patient to provide a measure of how and where the protons lose their energy. Several research groups are developing prototype systems based on different designs. In the current prototypes described in a recent review [7], the calorimetry and tracking are based on different technologies: For track reconstruction purposes, Scintillating Fibers or Silicon Strip Detectors are the most commonly used, which are based on one-dimensional strip readout in several rotated planes for tracking purposes. For the calorimetry part, crystal calorimeters such as CsI:Tl, YAG:Ce, NaI:Tl and plastic scintillator telescopes are commonly applied. One group has built a prototype with a (multi) layered CMOS pixel sensor telescope for calorimetry [8].

In this study, the feasibility of using a high-granularity Digital Tracking Calorimeter (DTC) for tracking and measuring the range and energy of individual protons in a proton beam is investigated through experiments and Monte Carlo (MC) simulations. The DTC consists of multiple layers of Monolithic Active Pixel Sensor chips with a digital readout, interleaved with a heavy material for energy absorption. The goal is to be able to register and separate a large amount of proton tracks in each data readout cycle: The requirement for a 10 s proton CT scan is 10 MHz [7]. While this study focuses on the calorimeter part of the setup, the sensor chips are considered near optimal for use in the tracking as well, due to their data processing capacity at the required readout speed, their high granularity and due to their short radiation length. The DTC was originally designed for the reconstruction of high-energy electromagnetic showers, and energy absorbers with high-density material was applied for that purpose. The large spacing of 32 mm Water Equivalent Thickness between the sensor layers is reflected in the final range resolution.

The production of a complete proton CT setup using the DTC is at present in the planning stage in collaboration between among others the University of Bergen, Bergen University College, Haukeland University Hospital and Utrecht University. A DTC further optimized for use in a proton CT system will utilize next-generation Monolithic Active Pixel Sensors with larger sensor areas readout speeds in the order of 5 µs, which may increase the rate capabilities towards the GHz rate. By optimizing the materials and geometry, the range resolution is expected to be limited by the range straggling.

In Section 2, the DTC is described. In Section 3 the setup for the beam measurements, which were performed at the Kernfysisch Versneller Instituut – Center for Advanced Radiation Technology (KVI-CART) in Groningen, the Netherlands, are described. In Section 4 the MC software and the key parameters in these are described. The data analysis is described in detail in Section 5. In Section 6 the results are presented, and, lastly this presentation ends with a discussion and the conclusions.

## 2   The Digital Tracking Calorimeter

A high-granularity digital sampling pixel detector is made available through participation in the ALICE-FoCal collaboration at CERN [9,10]. It is one of the proposed upgrades of the detector experiment carried out to provide an electromagnetic calorimeter for measurements of particle distributions at large rapidity $y$. The high pixel granularity allows for discrimination of $\gamma/\pi^0$ particles at very high momenta. The detector's small Molière radius of 11 mm enables that the electromagnetic showers originating from the high energy particles can be fully contained within the full calorimeter of 24 telescopic sensor layers, sandwiched between tungsten absorbers.

### 2.1   The MIMOSA23 chips

The Monolithic Active Pixel Sensor chip PHASE2/MIMOSA23 is applied in the prototype electronics assembly reported on in this paper. This sensor is a CMOS based digital high-granularity pixel sensor produced at the Institut Pluridiscipline Hubert Curien in Strasbourg in France [11]. The size of the active area is 19.2x19.2 mm$^2$, with a 640x640 array of 30x30 µm$^2$ pixels. The sensor's active epitaxial layer is 14-20 µm thick, and has a resistivity of either 10 Ω cm or 400 Ω cm. The chips were manufactured with different thicknesses and resistivities in order to assess and quantify their performance with these different parameters. The readout is 1-bit digital with a programmable signal threshold to adjust for electronic noise. The rolling shutter readout has a cycling time of 642 µs. More details about the specifications and performance of the MIMOSA23 chips can be found in [9,11].

Not all of the MIMOSA23 chips used in the experiment were working properly at the time. The result of this is that certain areas with bad sensors did not transfer information about which pixels that were activated, the so-called *hits*, and the analysis has to take this less than perfect efficiency into account accordingly by allowing for missing hits to occur in the analysis and reconstruction processes.

## 2.2 Geometry of the calorimeter

Two MIMOSA23 chips are mounted side-by-side to form a *module*. Such a module is shown in **Figure 1**. Two modules are put on top of each other, one rotated at 180 degrees with respect to the other, one facing the other, enabling that four sensor chips are placed at approximately the same depth in the longitudinal direction of the detector, i.e. along the central beam axis. There is a 100 µm gap between the two chips in a module, and when two modules are placed on top of each other there is a 90 µm overlap between the sensitive areas of each module. The lateral size of the sensitive area in a layer is 38.5x38.3 mm$^2$.

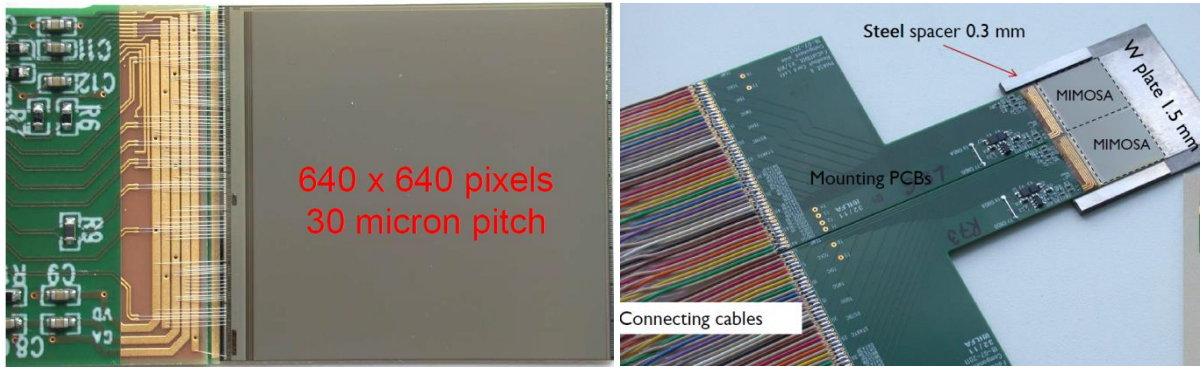

**Figure 1: Left:** *One MIMOSA23 chip connected to PCB.* **Right:** *Two MIMOSA23 chips mounted on PCB which is glued onto a tungsten absorber to form a module. The rainbow-colored readout cables are attached.*

Multiple sensor layers, 24 layers in total, are stacked behind each other, interleaved with 3.3 mm thick tungsten plates acting as an energy absorption material between the layers. A detailed description of the materials in a single layer is listed in **Table 1**. By using the formula $X_0^{-1} = \sum_i V_i/X_{0,i}^{-1}$ [12], where $V_i$ are the volume fractions of each material, the resulting radiation length $X_0$ was found to be 4.2 mm. The total thickness of a layer is 3.975 mm, thus the radiation length of a single layer is 0.97 $X_0$, or around 32 mm in units of Water Equivalent Thickness (WET). In the first layer, the absorbing material in the front end of the detector is a 0.02 $X_0$ thick aluminum plate. In this way, the beam is less degraded and scattered prior to reaching the first sensor layer, compared to the situation when applying tungsten.

| Material | Thickness [µm] | Radiation thickness | Density [g/cm$^3$] |
|---|---|---|---|
| W absorber | 1500 | 0.428 $X_0$ | 19.30 |
| Silver glue | 40 | 0.001 $X_0$ | 3.2 |
| PCB | 160 | 0.002 $X_0$ | 1.85 |
| Silver glue | 40 | 0.001 $X_0$ | 3.2 |
| MIMOSA23 | 120 | 0.005 $X_0$ | 2.33 |
| Air gap | 170 | 6E-06 $X_0$ | 0.001 |
| W absorber | 300 | 0.086 $X_0$ | 19.30 |
| Cyano-acrylate glue | 70 | 0.0002 $X_0$ | 1.0 |
| W absorber | 1500 | 0.428 $X_0$ | 19.30 |
| Air gap | 75 | 3E-06 $X_0$ | 0.001 |

**Table 1:** *The materials and their key properties, as used in the MC setup. The thicknesses are displayed both in terms of geometric thickness and the corresponding radiation thickness in units of the radiation length $X_0$.*

## 2.3 The complete detector setup

The calorimeter prototype is mounted onto a steel structure containing a system for liquid cooling of the electronics, mounting of the data readout electronics, as well as support for all the patch cables. The steel structure fixes three polyvinyl toluene scintillators to the system. The scintillators are used as suppliers for a trigger signal, and signals in coincidence from the scintillators trigger storage of an event. An image of the detector setup with all the layers and support structures is shown in **Figure 2**, and schematically also in **Figure 3**. A more detailed explanation about the setup and trigger logics can be found in [9,13].

### 2.3.1 Chip alignment correction

The layers need to be aligned in software before data analysis, in order to correct for misalignments from the fabrication process. Left unaligned, systematic lateral shifts would occur in the proton tracks between each layer, which would reduce the quality and efficiency of the track reconstruction. Position calibration has been performed at Utrecht University by aligning the tracks of cosmic muons; this has resulted in alignment correction values for each chip. The alignment correction defines lateral shifts and rotations of the chips, which has been applied on the experimental datasets.

## 2.4 Data readout

The patch cables lead to four 96-port Spartan FPGAs, which are further connected to two Virtex-6 FGPAs for triggering and multiplexing of the signal. For each readout cycle, 24 layers × 4 chips × 640 × 640 1-bit pixels are readout. This corresponds to a data size of 4.9 MB. The buffer size of the system is 4 GB, so in total 816 full readout cycles; called *frames*, can be read out in a single proton spill before the slower data transfer to the DAQ computer is performed. The 1-bit readout signal gives no information about the intensity of the detected signal from the traversing particles. A preset noise threshold determines if sufficient charge has been collected in each pixel for the pixel to register a signal. The numbering scheme of the chips is $\text{Chip} = 4 \cdot \text{Layer} + q$, where $q$ is the clockwise quadrant. Further details of the readout of this chip can be found in [13].

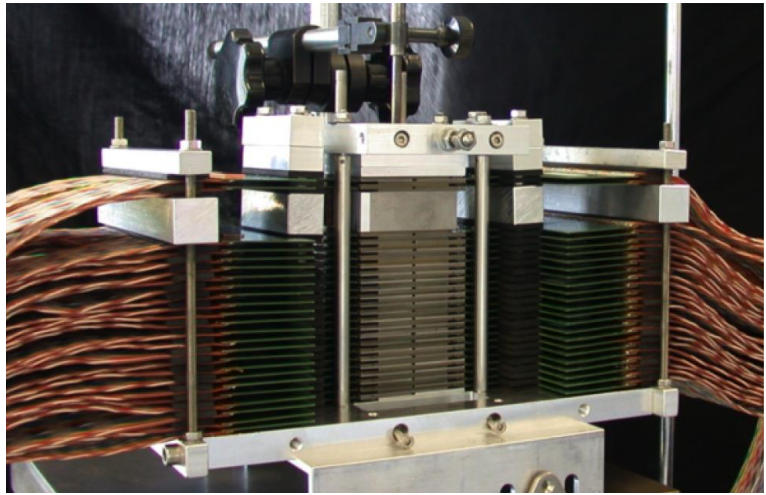

**Figure 2:** *The complete prototype detector setup. The 4x4 cm$^2$ sensor layers are not visible in the figure, these are located inside the central part of the structure. The modules; described in Section 2.2, are connected to the flat multi-colored readout cables. These are in turn connected to a patch-panel distribution unit, which facilitate the buffering and transmission of the 61 Gb/s signal to FPGAs through the patch cables* [9].

While the readout frequency is limited to 2 kHz, a higher effective readout frequency $f_\text{eff}$ can be achieved by accumulating $n_p$ protons in a single readout cycle: $f_\text{eff} = n_p \times 2 \text{ kHz}$. The high granularity of the detector results in a good ability for event separation. The granularity also determines the saturation limit $n_{p,\max}$. This saturation limit is found through the MC simulations and the result is then applied for the experimental dataset.

### 2.4.1 Threshold settings

The pixels in the MIMOSA23 chip are activated when the integrated charge has reached a preset threshold value. The electronic noise in MIMOSA23 has been estimated to be about 10 $e^-$ ENC [14]. The threshold is defined in units of the *fake rate*, or the probability of a pixel being activated by electronic noise. The fake rate settings used in the KVI beam test is $10^{-5}$, corresponding to 4 pixels per 640x640 chip. This fake rate is equivalent to a signal threshold of about 26 $e^-$ ENC, found by using a Poisson distribution with $\lambda = 10$ $e^-$ ENC. The choice of fake rate determines the charge diffusion cluster size described in Section 4.3. A separate study [15] using an analytical model from [16] to determine the diffuse cluster size has later confirmed the numbers outlined in this section.

## 3 Measurements in a proton beam

The measurements reported upon in this study were performed in December 2014 at KVI-CART in Groningen, the Netherlands. The cyclotron at the AGOR facility for Irradiations of Materials (AGORFIRM) delivers proton beams with energies from 40 MeV and up to 190 MeV [17].

### 3.1 Beam specifications

The sensors have a surface area of 4x4 cm$^2$, and during the experiment, the proton beam was shaped to this same field size. The intensity of the beam was set for delivering at most one proton per readout frame, with a detector readout frequency of $1/642$ μs $\simeq$ 2 kHz. The proton frequency is estimated to have been approximately 1.35 kHz, this value is deduced from the finding that about 67% of the readout frames contains proton tracks. See **Table 2** for a comparison between the number of readout frames and the number of reconstructed tracks.

| **Final energy** | 150 MeV | 160 MeV | 170 MeV | 180 MeV | 188 MeV |
|---|---|---|---|---|---|
| **Al degrader thickness** | 35 mm | 27 mm | 17 mm | 8 mm | 0 mm |
| **Number of readout frames** | 819 | 762 | 4944 | 1334 | 2739 |
| **Number of reconstructed tracks** | 408 | 408 | 3431 | 901 | 2010 |

**Table 2:** *List of beam energies applied at the KVI-CART beam test, the number of readout frames as well as the number of reconstructed proton tracks at each energy.*

The beam energies during the data acquisition were chosen with the motivation of applying the maximum available energy, and thus measuring the corresponding maximum proton range, in the multi-layered detector. Due to the high-Z absorber material, the 188 MeV proton beam is traversing through only the first 7 of the 24 layers: A beam energy of 450 MeV would have been needed in order for the protons to traverse the whole detector (all the 24 layers) in the longitudinal direction.

In order to deliver the different beam energies, the beam was degraded by the presence of an aluminum absorber in the beamline. For details about the applied proton beams; energies, as well as the different degrader thicknesses and the number of recorded protons in each of the setups, see **Table 2**. An energy spread of about 2-5 MeV is introduced from the degradation, this energy spread was calculated using GATE simulations. The energy spread increases with the thickness of the degrader. More detailed beam specifications and the beam optics are described in [17].

### 3.2 Data format and conversion

The raw data format from the experiment is a multiplexed data stream containing trigger information and the output of each of the 96 sensor chips. Additionally, pedestal runs with information about sensor noise are also available. These data streams are de-multiplexed and the pedestal noise is subtracted from each of the sensors. This data conversion was performed on the measurement data during a stay at the Utrecht University in January 2015, by using the conversion software developed at the Utrecht University [18]. The resulting output was stored as event-by-event objects containing the sensor layer number and pixel number for each of the activated pixels, by using the ROOT framework [19]. The output ROOT files were then applied in the analysis described in this work.

# 4 Monte Carlo simulation

A MC simulation was performed of the complete detector setup, as described in Section 2.3. The MC software GATE 7.0 [20] for Geant4 9.6.4 was used for this purpose. In addition to this, semi-empirical values for ranges and range straggling have been extracted from the PSTAR database [21].

## 4.1 Monte Carlo software setup

The MC software package GATE [20] is an application of the C++ based MC code Geant4 [22]. The GATE package was applied in the MC simulations as it simplifies the usage of Geant4, as well as adding features for simulations of detector functionality, such as the possibilities for simplified geometry building, readout logistics and triggering systems. The input to GATE is given through macro files which require no compilation before being executed.

### 4.1.1 Layer material properties

The prototype detector itself consists of 24 sensor layers, with the first differing from the rest by the use of aluminum as absorber material in front of the sensor. The geometry and materials in the layers as they were implemented in GATE are presented in **Table 1**.

A constant thickness for the epitaxial layer of 14 µm has been chosen, in order to simplify the simulation. The differences arising from the chips' thicknesses have been taken into account by calibration of the sensitivity of the charge diffusion process, as will be described in Section 4.3.1.

### 4.1.2 Geometry of the full setup

The geometry implemented in the simulations consists of three scintillators and the prototype detector. The geometry of the detector setup is described in Sections 2.3 and 4.1.1, and has been implemented in GATE accordingly.

Three polyvinyl toluene scintillators are used for signal triggering: A 1x4x0.5 cm$^3$ vertical scintillator, a 4x1x0.5 cm$^3$ horizontal scintillator and a 4x4x1 cm$^3$ front scintillator were placed 17.4 cm, 16.6 cm and 6.5 cm upstream of the front face of the detector, respectively. The detector setup including the scintillators as visualized in GATE is shown in **Figure 3**.

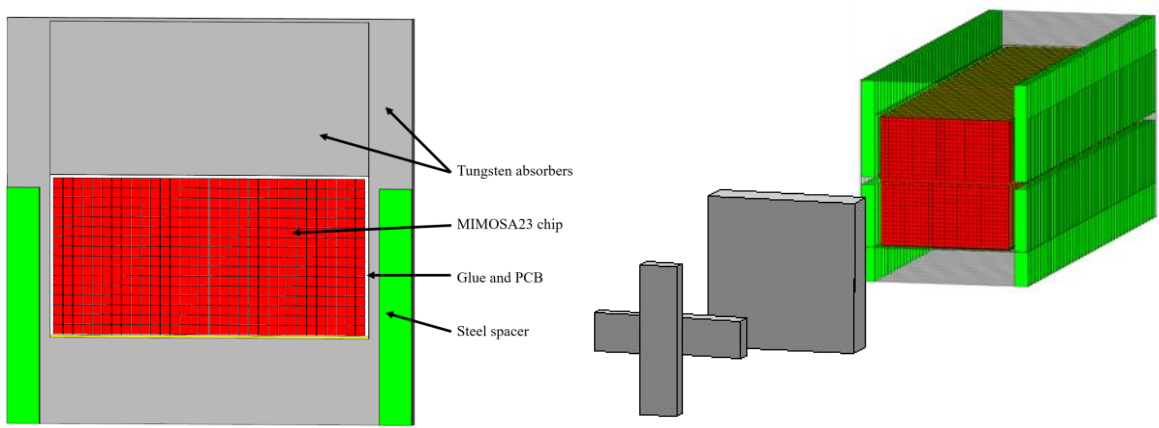

**Figure 3: Left**: *The MC implementation of a module. Two MIMOSA23 chips are mounted side-by-side, each with 640x640 pixels of 30x30 µm$^2$. The chips are mounted on PCB which is glued onto a tungsten absorber. Each sensor layer consists of two such modules internally rotated 180°, such that the two modules face each other. The area covered by the 2x2 MIMOSA23 chips is 38.5x38.3 mm.* **Right**: *The complete calorimeter consists of 24 sensor layers, also included are the three scintillators located upstream of the prototype detector.*

### 4.1.3 Physics builder lists

In GATE 7.0, the applied physics is chosen from a so-called physics builder list. Different physics builder lists are recommended for use in proton therapy and detector simulations. In [23] the recommendation is to use the *Standard Physics List Option 3*, which has a special emphasis on increased simulation accuracy around the Bragg Peak area. In order to include nuclear interaction processes, the physics list *QGSP_BIC_EMY* is used. The values for the production thresholds are set to 0.1 mm, i.e. a new particle from a decay or collision is required to have an energy corresponding to at least this range in order for the particle to be transported further, otherwise the energy is deposited locally. The minimum proton step size is set to 0.05 mm. Since *Option 3* is used, the minimum step size decreases towards the Bragg Peak depth. The adjustable mean excitation potential $I$ for water is set to 75 eV to match the value used in the PSTAR data tables [21].

### 4.1.4 Beam modelling by the use of a General Particle source

The proton beam at KVI was described in Section 3.1. The beam is represented by a source in the GATE simulations. The source is defined as a General Particle Source-based mono-energetic beam using the nominal energies of the beam test. The protons are emitted from a square 4x4 cm$^2$ plane, with no angular deflection.

### 4.1.5 Monte Carlo information through scoring

The scoring defines the variables that are stored during each MC simulation. Every interaction with energy deposition in a Sensitive Volume (SV) is written to an output file. The 96 separate MIMOSA23 chips, each with a thickness of 14 µm, are defined as separate SVs. If the 41 million pixels had been defined as individual SVs, a rather large number of volumes would have had to be stored in the memory.

## 4.2 Range calculation

The projected proton range was obtained by comparison between GATE simulations and the PSTAR database [21]. A simple model has been proposed in [24] for the description of the relationship between energy and range for a beam of protons traversing into matter:

$$R_0 = a_1 E_0 \left[ 1 + \sum_{k=1}^{N=2} (b_k - b_k e^{-g_k E_0}) \right]. \tag{4.1}$$

$R_0$ is the projected range of the proton beam and $E_0$ is the initial proton energy. The dimensionless parameters $a_1, b_k$ and $g_k$ in MeV$^{-1}$, as well as the choice of $N = 2$, are determined through fitting of the model to range-energy data. An example of values obtained through a fit to nuclear data from ICRU49 [25] are tabulated in [24].

These parameters have been found both for water and for the calorimeter geometry, yielding an accurate method to calculate the Water Equivalent Thickness (WET) based upon knowledge of the range in the detector and of the beam energy. To calculate the energy at depth $z$ from the range, an inversion formula is derived in [24]:

$$E(z) = (R_0 - z) \sum_{i=1}^{N=5} c_k e^{-\lambda_k (R_0 - z)}. \tag{4.2}$$

Where the coefficients $c_k$ and $\lambda_k$ are found by curve fitting of the model to range-energy data.

### 4.2.1 Error estimation of the range due to the sampling spacing

The error in the range measurements is dominated by the large sampling spacing between the sensor layers. Assuming uniform hit distributions throughout each layer, the error is defined as $\Delta z/\sqrt{12} = 3.975$ mm$/\sqrt{12} = 1.15$ mm, corresponding to 10 mm WET. This error is visualized as the horizontal error bands in **Figure 5**. While this error is not propagated further in the analysis, it is connected to the error in the reconstructed range $\langle\hat{\sigma}\rangle$ as defined in Section 5.5.

## 4.3 Charge diffusion model

Large clusters with sizes varying normally between 3 and 35 pixels are activated by the charge diffusion of electron-hole pairs, which are liberated in the epitaxial layer of the sensor chip by a passing proton. This effect cannot be modelled straightforwardly in MC simulations due to the low electron energies associated with the charge diffusion process. The sizes of the clusters are roughly proportional to the proton's energy deposition, and the shapes of the clusters are approximately Gaussian distributed. The main procedure in the algorithm is thus to convolute the 2D array of MC simulated hits with a Gaussian function, in which the $\sigma$ parameter is depending on the deposited energy.

In order to include statistical uncertainties, the following phenomenological method was developed: For each of the hits, a Gaussian distribution surrounding the hit position is generated. The standard deviation of the Gaussian is set to $(\alpha E_{\text{dep}})^\beta$. The Gaussian will then be sampled $N$ times in order to by this introduce statistical uncertainties, and thus $n$ pixels are set as activated surrounding the original pixel, where $n$ is the number of unique pixels from the $N$ sampled pixels. By matching the cluster size distributions arising from the above model with experimental data, the values $\alpha = 0.24 \text{ keV}^{-1}$ and $\beta = 0.35$ were obtained. In the same way by applying $N = E_{\text{dep}} \cdot 2.7 \text{ keV}^{-1}$, one ensured that the shapes of the generated clusters would have a distribution corresponding to the average sizes of the measured clusters in the experimental data.

The model described above was used to create a parameterization for the estimation of the amount of deposited energy, given in keV, from the number of activated pixels, $n$, in a cluster. A polynomial fit to a large number of modelled clusters was used for this purpose:

$$E_{\text{dep}} = f(\text{Chip})[-4.0 + 3.88\, n + 1.24 \cdot 10^{-2}\, n^2 - 1.14 \cdot 10^{-3}\, n^3 - 1.42 \cdot 10^{-6}\, n^4]. \quad (4.3)$$

The deposited energy given in units of keV/μm can be found by dividing $E_{\text{dep}}$ by the thickness of the sensitive epitaxial layer, which has been modelled to be 14 μm. The scaling factor $f(\text{Chip})$ is explained in the following section.

### 4.3.1 Chip sensitivity calibration

While the charge diffusion model described in the previous section can be used to estimate the $E_{\text{dep}}$ from the cluster sizes obtained from both MC simulations and the experimental data, the physical chips exhibit thickness and sensitivity variations. It follows that this model describes only the average chip sensitivity. Calibration of the sensitivity of each chip, viz. a scaling factor in the $E_{\text{dep}}$ calculation, ensures that the responses of the physical chips are uniform throughout the whole calorimeter. The calibration is performed by finding a scaling factor $f(\text{Chip})$ applied to Eq. (4.3) for each physical chip. In Section 6.2 the results from this procedure are presented.

### 4.3.2 Error estimation from the charge diffusion model

The error associated with the model described in the Section 4.3 is assumed to be the random error from the number of activated pixels, given by $\sigma_n = \sqrt{n}$. By propagating the error, we arrive at the following:

$$\sigma_{E_{\text{dep}}} = \sigma_n \frac{dE_{\text{dep}}}{dn} = \sqrt{n} \cdot f(\text{Chip}) \cdot (3.88 + 2.48\, n - 3.42 \cdot 10^{-3}\, n^2 + 5.68 \cdot 10^{-5}\, n^3). \quad (4.4)$$

# 5 Data analysis

The data analysis was performed on the data files containing event-by-event objects, as described in Section 3.2. The analysis, including the tracking algorithm and resulting energy estimation, has been written in the ROOT framework by applying C++ programming code [19].

## 5.1 Noise

There is a certain amount of unavoidable noise present in the beam test data. Prior to each change of the beam parameters, as well as during the run, pedestal values were read out in order to calibrate the noise level of the individual pixels. The process of pedestal removal on the MIMOSA23 chips is described in [16], and this had already been performed on the experimental data by provided the Utrecht University at the time of data retrieval for our present use of the data files.

The clusters generated by the proton tracks typically activate a quite large pixel area due to the charge diffusion processes, as described in Section 4.3. Therefore, it is quite straightforward to remove all the remaining noise, which normally appears as isolated one- and two-pixel clusters.

## 5.2 Clustering

Since each traversing proton activates a cluster of pixels through charge diffusion process, these clusters must be identified. All activated pixels, called *hits*, surrounding a proton track should be incorporated into a single cluster. This is done through a simple neighboring algorithm: For each hit, check if any of the eight possible neighboring pixels have been activated. This algorithm is then run recursively on all the activated neighboring pixels. The resulting cluster, with its center-of-mass position and number of pixels, is then stored.

## 5.3 Tracking

The individual clusters in each layer are connected through proton tracks. A track-finding algorithm has thus been developed, modelled after the *track following* procedure in [26]: Cluster pairs at approximately the same lateral position in the two first layers are identified. Multiple cluster pairs may originate from the same seed cluster in the first layer, due to tracks having different initial vectors. Using the position and direction of each cluster pair as the starting point of a growing track, further clusters are searched for at extrapolated anticipated positions in deeper layers. At each sequential layer, a search cone is applied in order to identify all possible matches. Within this search cone, the cluster that is closest to the anticipated position is added to the growing track. The radius of the search cone is calculated as the $k \cdot \theta_0$ value of the expected Multiple Coulomb Scattering (MCS) angle distribution with a standard deviation $\theta_0$. The value $k$ is chosen, and $\theta_0$ is found by [27]:

$$\langle \theta_0 \rangle = \frac{13.6 \text{ MeV}}{\beta pc} \sqrt{d/X_0}\,[1 + 0.038 \ln d/X_0], \tag{5.1}$$

where $\beta$ is the relativistic speed $v/c$, $pc$ is the momentum in MeV, $d$ is the thickness of the layer and $X_0$ is the average radiation length of the calorimeter, which was as presented in Section 4.1.1 as 4.2 mm. The factor $\beta pc$ is found from the calculated remaining proton energy in a layer, as defined in Eq. (4.2), and the expected MCS angle $\langle \theta_0 \rangle$ is found for each layer, given in radians.

All candidate tracks originating from the same seed cluster in the first layer are compared, by searching for the highest scoring track. The track score is calculated as a function of the track length, the amount of angular scattering between each layer, and a check of whether the track contains a Bragg Peak or not. The tracking algorithm is run twice, first with $k = 2.5$ and then with $k = 5$, so that first the relatively straight and most abundant $\theta < 2.5\,\theta_0$ tracks are found, and then to ensure that all tracks with $2.5\,\theta_0 < \theta < 5\,\theta_0$ are found. It turns out that a number of tracks are still incorrectly reconstructed by comparison with MC simulations, which is assumed to be the ground truth.

### 5.3.1 Track reconstruction quality

Some track optimization methods are performed, in order to ensure that each proton track is reconstructed with high accuracy.

Due to the physics of proton interactions, a portion of the protons will stop abruptly prior to entering their Bragg Peak region, in which the bulk of the protons come to rest due to nuclear interactions. Such tracks do not contain a Bragg Peak, and thus do not have an increased energy deposition in the deepest layer. A cut based on the $dE/dx$ value of the deepest cluster of each reconstructed proton track ensures that such tracks are identified and removed from the analysis. A minimum value of 3 keV/μm has been chosen based on the expected $dE/dx$ values around the Bragg Peak area, see **Figure 4**. Tracks that leave the detector laterally, and thus do not contain a Bragg Peak, are also removed from the analysis by identifying outwards-pointing tracks that end near a detector side.

As mentioned in Section 2.1, some of the sensor chips did not transfer all of the hits to the DAQ system. In addition, there is 100 μm dead area between the two chips in a module. The track reconstruction software needs to allow for missing clusters due to cluster merging from overlapping or sparsely separated proton tracks, dead chips, bad data channels, etc. If the reconstruction algorithm cannot find a cluster to append to the growing track, it extrapolates the track one layer further, based upon the track position in the last layer with an identified cluster. If there is a cluster close to the extrapolated position in this next layer, the track continues from there. In this way, tracks that are lacking a cluster in a single layer are reconstructed. The survival rate of the protons is lower in the experimental dataset compared to MC results: Areas with few reconstructed tracks near the projected position of the bad chips can be observed.

Due to charge diffusion, densely separated hits may be identified as a single merged cluster. Therefore, a cluster splitting algorithm is applied. Two crossing protons can produce a single merged cluster with the result that only one of the reconstructed tracks may incorporate it, resulting in a track with a missing cluster in the layer where the protons crossed. The cluster splitting algorithm locates all the crossing track pairs in the layer where one of them is missing a cluster. It then divides the supposedly merged cluster into two halves, and connects the new cluster to the track without a cluster. Each of the new clusters has a smaller size than that of the merged cluster: The size is chosen according to Eq. (4.3) such that the total amount of deposited energy is conserved.

### 5.3.2 The effects by track multiplicity on saturation

During the beam test, the frequency of the beam was matched to the readout frequency of the calorimeter, which yielded at maximum one proton in each readout. Due to the high granularity of the MIMOSA23 sensors, it is possible to reconstruct a large number of proton tracks concurrently. Considering this, $n_p$ readout frames have been accumulated in the track reconstruction step.

The detector occupancy is the ratio of the number of activated pixels to the total number of pixels in the detector. The detector occupancy increases linearly with $n_p$, and at $n_p = 500$ the detector occupancy in the layer containing the Bragg Peak is 0.42%. This corresponds to 13.5 activated pixels per proton track in that layer.

A higher detector occupancy decreases the probability that all hits in a given reconstructed track originates from the same primary proton. The number of correctly reconstructed tracks has been found through checks against the event ID from MC simulations. The saturation limit $n_{p,\max}$ will be determined based on the applied tracking algorithm, estimating the maximum effective readout frequency $f_{\text{eff}}$ using MC simulations.

## 5.4 Range fitting

Each track contains a number of clusters and there is maximum one cluster in each layer for each proton track. The proton range is converted into WET using the method described in Section 4.2. The cluster sizes are converted into estimated deposited energy using the method described in Section 4.3.

By fitting a Bragg Curve model to the deposited energy in each layer, the estimated proton range $\widehat{R_0}$ can be calculated with an improved accuracy compared to that of using the last traversed sensor layer in the track as representing the range of that proton. Although using Eq. (4.1) may be accurate enough for calculation of the range based upon information about the initial energy, a simpler function such as the differentiated Bragg-Kleeman rule [28] has a smoother behavior around the Bragg Peak and is as such more suitable for this purpose:

$$D_{\mathrm{BP}}(z) = -\rho^{-1}\frac{\mathrm{d}E}{\mathrm{d}d} = \frac{1}{\rho p \alpha^{1/p}(R_0 - z)^{1-1/p}}. \tag{5.2}$$

Here, $z$ is the traversed depth, $R_0$ is found using Eq. (4.1) and is used as input for the model, $\rho = 1\ \mathrm{g/cm^3}$ for pure water and the parameters $p$ and $\alpha$ are found in a similar fashion to those described in Section 4.2. By fits to PSTAR data in the energy range between 150 and 250 MeV one obtains the values $\alpha = 0.0446\ \mathrm{cm/MeV}$ and $p = 1.668$. Eq. (5.2) exhibits range errors at large $(R_0 - z)$ values due to the underlying simplifications [28], the range estimation is however quite accurate near the Bragg Peak where $R_0 = z$.

It should be noted that the range estimation takes also into account the energy loss of the incident protons in the scintillators providing trigger signal (see Section 4.1.1). The number of traversed scintillators is estimated from the initial proton vector, and a pre-sensor WET value is added to each proton track according to the estimated energy loss.

## 5.5 Estimating the range from multiple proton tracks

By performing the range fitting procedure as described in Section 5.4, an estimate of the range $\widehat{R_0}$ and of the corresponding initial energy $\widehat{E_0}$ are obtained for each reconstructed proton track. The validity of this estimate depends to a large degree on the position of the Bragg Peak relative to the position of the sensor layer where the proton comes to rest. In a longitudinally segmented detector setup, depending on whether the energy of the protons is such that the Bragg Peak depth is located within one sensor layer, or if the Bragg Peak depth is located between two sensor layers, different results are obtained. If the Bragg Peak is located within a sensor layer, the resulting distribution of reconstructed ranges is normally distributed with a central value at the sensor layer depth. By increasing the beam energy slightly, the distal tail of the $\widehat{R_0}$ distribution reaches into the next, deeper, sensor layer. Two separate normal distribution appear, each representing the central position of two sensor layers most adjacent to the center value of the physical Bragg Peak.

A Gaussian fit is performed around the depth of each sensor layer in order to identify the distributions of the $\widehat{R_0}$ values. Each fit is evaluated based on the sum $n$ of the bin values in the $\mu \pm 3\sigma$ region, as well as its $\chi^2/n$ value. The rejected fits are usually positioned in areas with high noise or low statistics, and the cut criteria have been chosen as $n < 0.2\ N$ and $\chi^2/n > 8$. Here, $N$ is the total number of entries in the histogram.

The resulting Gaussian distributions are determined by $(\mu_1, \sigma_2)$ and potentially $(\mu_2, \sigma_2)$ if a second distribution is found, where $\mu_1 < \mu_2$. This procedure is performed in order to find the histogram bin with the lowest range value $x_{i'}$ in the Bragg Peak region, having bin height $w_{i'}$. The range value for the bin is defined relative to the first Gaussian, where $x_{i'} = \mu_1 - 3\sigma_1$. As noted in Section 5.3.1, some reconstructed protons tracks stop abruptly due to nuclear interactions, and all tracks identified as stopping due to nuclear interactions are removed from the analysis. This cut on range will further ensure that the overall range estimate is based only on proton tracking stopping in a Bragg Peak.

The mean value of all range estimates in the Bragg Peak region of the dataset is defined as the reconstructed range $\langle\widehat{R_0}\rangle$, and the corresponding standard deviation is $\langle\hat{\sigma}\rangle$. These parameters are estimated by the sums in Eq. (5.3).

$$\langle \widehat{R_0} \rangle = \frac{\sum_{i=i'}^{\infty} w_i x_i}{\sum_{i=i'}^{\infty} w_i}, \qquad \langle \hat{\sigma} \rangle = \sqrt{\frac{\sum_{i=i'}^{\infty} w_i (x_i - \langle \widehat{R_0} \rangle)^2}{[\sum_{i=i'}^{\infty} w_i] - 1}}. \tag{5.3}$$

By calculating the standard deviation of the reconstructed range, without making any assumptions of the shape of the distribution of the data or by performing any fits to the data, the range resolution $\langle \hat{\sigma} \rangle$ for a given dataset can be estimated in a direct way. In this way, one avoids the propagation of the errors connected a mean value of multiple Gaussian fits.

The accuracy of the reconstructed range $\langle \widehat{R_0} \rangle$, expressed through the standard deviation $\langle \hat{\sigma} \rangle$, is limited to be larger than the range straggling originating from the individual interaction histories that each of the protons in a proton beam undergo. This lower limit of the accuracy can be estimated to be $\langle \hat{\sigma} \rangle_{\min} = 0.017 \, R_0^{0.935}$, in units of WET. This is the range straggling as it observed in MC simulations where the full detector volume has been scored, including the energy absorbers. This value is slightly larger than the range straggling occurring in pure water, which is approximately $0.012 \, R_0^{0.935}$ [29].

# 6 Results

In the following section the results from the beam tests are presented, and these results are compared to the results from the MC simulations.

## 6.1 Monte Carlo validation

The complete setup has been simulated in the GATE application of Geant4, and the results from many different simulations have been applied when developing the analysis framework. The same analysis methodology is applied to results from both MC simulations and experiments, the only differences being the readout format and whether or not the charge diffusion model described in Section 4.3 should be applied.

The results from the MC modelling is compared to the experimental data through the reconstructed ranges as described in Section 5.5. This is shown for a single initial energy in **Figure 6**, and for all the different energies in **Figure 7**. While the experimental data exhibits a higher noise level in the range distributions, the general features are well represented through the MC simulation.

## 6.2 Accuracy of the charge diffusion model

A model for the charge diffusion around a proton track was presented in Section 4.3. A comparison between the model and the experimental data is presented in **Figure 4**, where the experimental data represents the measured values of the size of the charge diffused clusters, and the charge diffusion model is applied on the MC dataset. The different datasets used to this end were described in Section 3.1. In both cases, the cluster sizes have been converted into units of keV/µm in order to account for the chip sensitivity calibration.

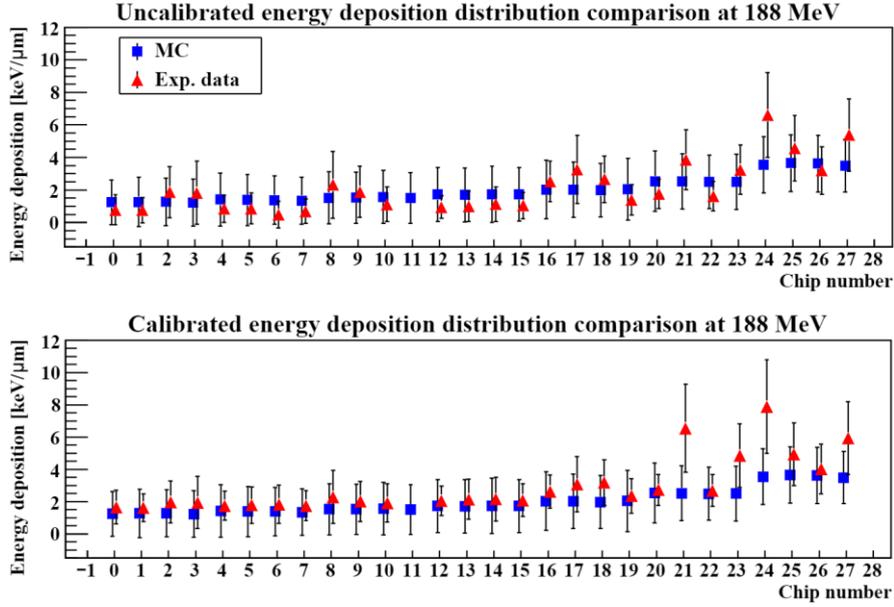

**Figure 4:** *The estimated energy deposition for each chip, displayed for purposes of validation of the charge diffusion model applied on the MC dataset and of validation of the chip sensitivity calibration on the experimental data. The result of the charge diffusion model is compared to the experimental data, in which the sizes of the charge diffused clusters are measured directly. In both cases Eq. (4.3) is applied to calculate the energy deposition. The data points correspond to the mean value of the distribution of the energy deposition, while the error bars correspond to the RMS value of the distribution.*

The MC charge diffusion model has been calibrated to the *average* cluster sizes in the experimental dataset. The sensitivity calibration of the sensors has been performed through finding a unique scaling factor $f(\text{Chip})$ for each of the 28 chips, corresponding to 7 layers, for which there exists data. In the first few sensor layers, the variation of energy deposition is small, so that the scaling factors for a given layer exhibit low variation between the datasets of different energies. In the Bragg Peak regions of the different energies, the energy deposition variation is high, such that there is less agreement between the scaling factors of the different energies. In **Figure 4**, this is shown through a quite good agreement below chip number 16, and less good agreement above.

## 6.3 Accuracy of the range estimation

After performing the range fitting as described in Section 5.4, estimates of the range $\widehat{R_0}$ and energy $\hat{E}$ is obtained for each individual proton track. The Bragg Curve fit to three individual protons tracks are displayed in **Figure 5**.

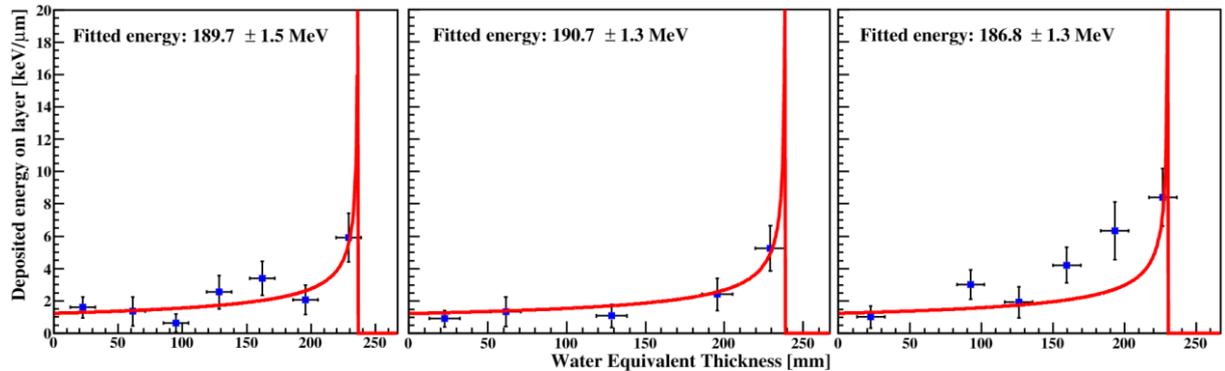

**Figure 5:** *Bragg Curve fit to measurements from the 188 MeV experimental beam. Each plot represents the track from an individual proton from the experimental data, the data points display the total projected path length and the deposited energy in each traversed layer. The solid line is the*

*Bragg Curve described in Eq. (5.2). Note that in the middle and right panel, the track is missing measurements in the some of the layers. This is due to bad readout channels as discussed in Section 5.3. The tracks survive while skipping a sensor layer.*

As described in Section 5.5, the range distribution $\widehat{R_0}$ from multiple proton events is approximately Gaussian distributed around each sensor layer. A Gaussian fit procedure is applied in order to estimate the reconstructed range $\langle\widehat{R_0}\rangle$ of protons with different individual ranges $\widehat{R_0}$. Three examples of such Gaussian fits are shown in **Figure 6**, displaying both MC simulations results and experimental data. The agreement between the applied beam energy and the reconstructed energy is good in the 188 MeV case: $188 \pm 3$ MeV for MC simulations and $187 \pm 3$ MeV for the measurement. In the 170 MeV case, the depth of the Bragg Peak is split between two sensor layers, and the reconstructed energy based on data from the measurements is $167 \pm 9$ MeV.

A set of MC simulations was performed with beam energies varying from 145 MeV to 200 MeV, increasing the beam energy in steps of 1 MeV. The reconstructed ranges $\langle\widehat{R_0}\rangle$ from each dataset are shown in **Figure 7**. In the same figure, results from the analysis performed on the experimental data are also shown. An oscillatory pattern is observed in the reconstructed ranges. This pattern is due to the undersampling of the Bragg Curve, where the WET between the sensor layers is approximately 32 mm. The estimated mean range straggling in this energy range is 2.4 mm WET, which is a factor 4 below the observed range uncertainties.

### 6.3.1 Monte Carlo simulations

For beam energies between 145 MeV and 200 MeV, the average estimation error $\langle\hat{\sigma}\rangle$ based on Eq. (5.3) is 9.4 mm WET (4.6%), and the average absolute deviation from the expected range $\langle|R_0 - \langle\widehat{R_0}\rangle|\rangle$ is 3.3 mm WET (1.7%). The values for $R_0 - \langle\widehat{R_0}\rangle$ vary from -7.9 mm WET to 0.3 mm WET due to the oscillatory behavior of the estimation.

### 6.3.2 Experimental data

The experimental dataset consists of the energies as listed in **Table 2**. The average estimation error $\langle\hat{\sigma}\rangle$ from the complete dataset is 8.4 mm WET (4.1%), and the average absolute deviation from the expected range $\langle|R_0 - \langle\widehat{R_0}\rangle|\rangle$ is 8.2 mm WET (4.1%).

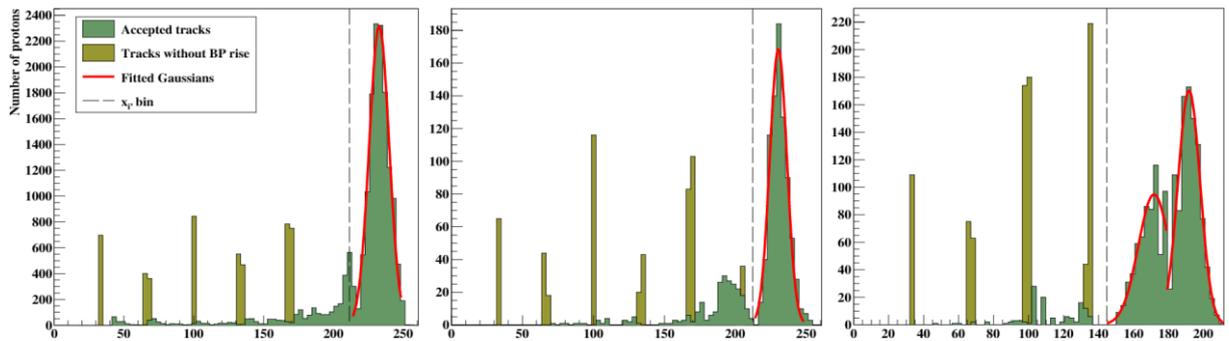

**Figure 6:** *Range estimation $\widehat{R_0}$ distributions from Bragg Curve model fits. The dashed line indicates the lowest-range bin $x_{i'}$ used for the range estimation, as described in Section 5.5. Note the short ranges due to nuclear interaction processes, identified due to the low $dE/dx$-values in the layer with the last recorded hit, these are shown as peaks at the sensor layer positions using a different color. The reconstructed energies $\langle\widehat{E_0}\rangle$ are* **Left:** *$188 \pm 3$ MeV from a 188 MeV MC simulated mono-energetic beam.* **Middle:** *$187 \pm 3$ MeV from the 188 MeV beam taken during the KVI-CART beam test.* **Right:** *$167 \pm 9$ MeV from the 170 MeV beam taken during the KVI-CART beam test.*

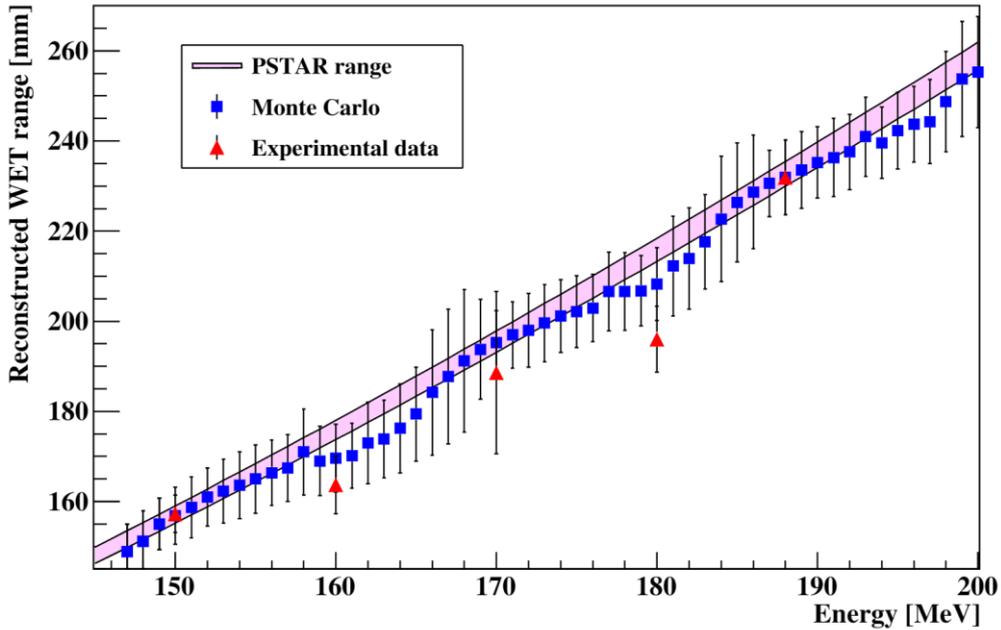

**Figure 7:** *Reconstructed ranges $\langle \widehat{R_0} \rangle$ of proton beams with different energies. Results from both the MC simulations and from the experimental measurements are displayed on the plot. The PSTAR range is displayed using a band representing the expected range straggling. Average numbers for the deviations between $R_0$ and $\langle \widehat{R_0} \rangle$ as well as the corresponding resolution $\langle \hat{\sigma} \rangle$ are presented in Section 6.3.*

## 6.4 Survival rate of tracks

Track loss may occur due to protons leaving the detector geometry, due to dead sensor areas and due to inelastic nuclear interactions in the detector. Tracks that are incorrectly reconstructed due to mismatch errors will be discussed in Section 6.5.

The overall reconstruction efficiency using all recorded beam energies is 60%. In this number, all identified inelastic nuclear interactions have been subtracted since they do not contribute to the range calculation.

### 6.4.1 Track loss due to nuclear interactions

A fraction of the tracks end before their expected range, this is mainly due to nuclear interactions. About 33% of the tracks stop prior to their energy dependent mean projected range, together with having a cluster size distribution with no identifiable Bragg Peak. The results are higher than the fraction of protons undergoing nuclear interactions obtained from [30] and from MC simulations, where the values across the energy range is found to be about 19%. The discrepancy may arise from tracks that resemble inelastic scattering, but are instead incorrectly reconstructed tracks from protons undergoing other processes.

### 6.4.2 Track loss due to dead sensor areas

If a stopping down track is assumed to stop in a dead sensor area, it is discarded. The fraction of tracks discarded in this manner is 5% across the detector area and beam energies.

### 6.4.3 Track loss due to protons leaving the detector

Some of the protons leave one of the lateral detector sides. This is either due to multiple scattering processes or due to the proton's initial direction when entering the front face of the detector. Protons leaving the detector before coming to rest are removed from the analysis. This effect is proportional to the proton range, and in average 4% of the protons exit from the detector this way.

## 6.5 Effective readout speed and reconstruction accuracy

The number of particles analyzed concurrently determines the effective readout speed. As described in Section 5.3.2 the saturation limit of the detector is given by $n_{p,\,\text{max}}$. This number is calculated according to the accuracy goal of the proton tracking. **Figure 8** shows the relationship between $n_p$ and the accuracy of the tracking algorithm. Tracks not surviving the effects described in **Section 6.4** are not included in this figure.

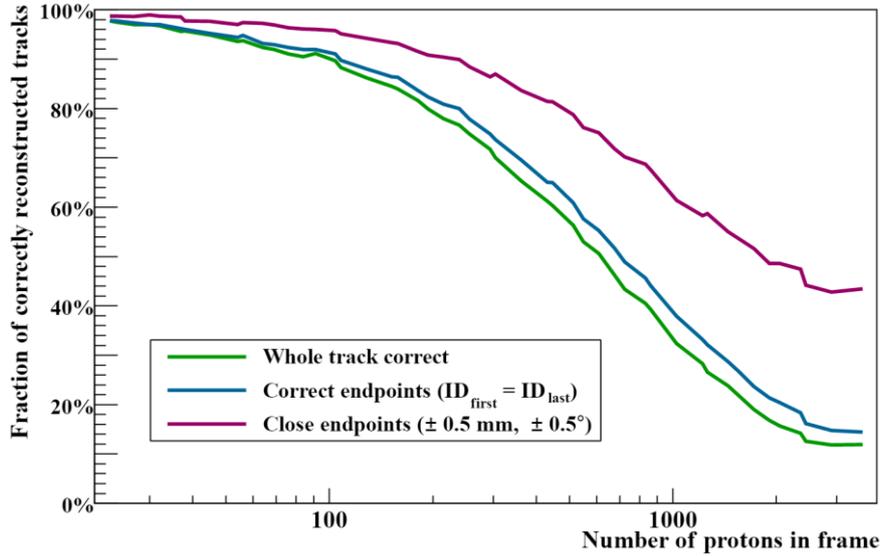

**Figure 8:** *The fraction of correctly reconstructed tracks, benchmarked using MC simulations. The track consists of one cluster in each layer, each cluster is tagged with an event ID in GATE. The figure shows the portion of tracks which are identified as correctly reconstructed, viz. they originate from the same primary particle. Three curves are shown according to different definitions of a correct track reconstructed, see Section 6.4 for details.*

The resulting saturation limit in which 80% of the tracks are correctly reconstructed is $n_{p,\,\text{max}} = 235$, resulting in an effective readout frequency of $f_{\text{eff}} = 470$ kHz. However, the misidentification of a track introduces potentially small errors, such as a small shift in its angular orientation and lateral position determination. With an allowance of small deviations; applying the values $\pm 0.5°$ and $\pm 0.5$ mm on the misidentified track in the first layer, the resulting saturation limit at 80 % accuracy is $n_{p,\,\text{max}} = 480$, corresponding to $f_{\text{eff}} = 960$ kHz, or 60 000 protons per cm$^2$ per s.

The results in **Figure 6** and **Figure 7** as well as the resulting range resolutions are obtained while applying $n_p = 500$. Note that while the actual proton frequency in the experimental proton beam was approximately 1.35 kHz, the saturation limit is found using MC simulations where each virtual readout frame contains a single proton. Due to this the saturation limit of this setup is $f_{\text{eff}} \simeq 1$ MHz.

## 7 Discussion

In this study, a number of specifications and properties of the DTC and of the corresponding MC modelling, as well as results obtained through MC simulations and experimental data have been presented. As this proof-of-concept calorimeter prototype was originally designed for application in a high energy regime, the choice of materials and of the geometrical layout is not optimal for use in a therapeutic proton beam. However, the results presented in this work indicate that the DTC is a detector concept with a promising potential in the proton CT context.

The geometrical sampling spacing between the active sensor layers is 32 mm WET. The results of those design choices are reflected in the range resolution of 8.4 mm WET (4.1%). The results show systematic errors of up to 8 mm WET range deviation, depending on the initial energy. One of the differences between the MC simulations and experimental data may be explained by the somewhat lower statistics in the latter. Other factors such as the beam quality and potentially unknown detector characteristics may also affect the interpretation of the acquired data. The range estimation error is dependent on the initial energy, displaying an oscillating pattern caused by the somewhat large sampling spacing.

This prototype needs further improvements in order to meet the requirement for range resolution of around 1% in a proton CT [7]. This requirement for range resolution is due to the range straggling limit of about 1%, which has been reached in [31]. Several other prototype scanners have a range resolution of 2-3% [32–34].

The effective readout speed in the order of 1 MHz, as presented in Section 6.5, depends on the quality of the tracking algorithm and its parameters. An improved algorithm would increase the accuracy of the track reconstruction, and allow for separation of a higher number of protons in each readout frame. The effective readout speed presented in this work is at the high end of the readout speeds of existing prototypes [7], where the current fastest proton CT systems exhibit readout speeds at 2 MHz [31,35]. In order to deliver a 10 second proton CT scan, a readout speed of 10 MHz is required [7].

The Bergen Proton CT research group is currently conducting research with the aim of developing the next prototype of a DTC, utilizing the potential of next-generation Monolithic Active Pixels Sensors. With readout speeds in the order of 5 µs and with larger sensor areas, the readout frequency could be increased to the GHz range. Utilizing low-Z absorber materials and optimized geometries, the range resolution is expected to be improved towards the range straggling limit, which is required for proton CT [7] purposes.

Another potential application for the next prototype of the DTC is its usage in combination with laser accelerated protons (LAP) [36]. While no clinical implementation of LAP has been shown, several feasibility studies of beam delivery [37] and treatment quality [38] are available. LAP is in its principle expected to deliver protons of therapeutic energies in very short picosecond bursts with kHz repetition rates. The DTC might be able applicable in resolving the resulting bursts of a few thousand protons per readout cycle by exploiting the high-granularity of the sensors which allows simultaneous tracking of individual protons as presented in this work.

# 8 Conclusion

In this study, the feasibility of applying a proof-of-concept version of the Digital Tracking Calorimeter (DTC) in a proton CT system has been shown. Methods have been developed for the purpose of calculating the energy deposited by protons, by modelling of the charge diffusion process of electron-hole pairs liberated by traversing protons in digital pixel sensors; for performing the subsequent track reconstruction through multiple sensor layers separated by energy absorbers; and for reconstruction of the initial energy of the proton tracks through the fitting of Bragg Curve models. The above methods have been presented and evaluated, using results from both Monte Carlo simulations and experimental measurements.

The results of this work indicate that the DTC can be used for track reconstruction and range estimation for a significant number of concurrent proton tracks at therapeutic energies. The materials used in the current version of the DTC are optimized for applications in a high energy physics experiment. Due to this, every sensor layer is separated by tungsten absorbers of Water Equivalent Thickness (WET) of 32 mm. This sets an upper limit to the accuracy with which the range and energy of protons can be determined in the present prototype.

The WET range of individual protons can be determined with a resolution of 4%. The required range accuracy in a proton CT setup is usually defined as the range straggling limit at about 1%, and in relation to this demand, the proof-of-concept DTC needs further improvements in order to meet this requirement. A high effective readout speed capacity of 1 MHz has been demonstrated, which is at the high end of the readout speeds of existing prototypes.

The results from this proof-of-concept tracking calorimeter shows that a next version with a more optimized prototype has the potential of enabling fast and accurate determination of the ranges of individual protons in a therapeutic proton beam.

# 9 Acknowledgements

This project was supported by Helse Vest RHF (Western Norway Regional Health Authority, Stavanger, Norway) grant [911933]. The project was co-financed by the European Union within the Seventh Framework Programme through IA-ENSAR (contract no. RII3-CT-2010-262010). The authors would like to extend thanks to Kristian Austreim and Ganesh Tambave at the University of Bergen and Gert-Jan Nooren at the Utrecht group for the KVI-CART beam test data and details about the calorimeter, to Liv Bolstad Hysing at Haukeland University Hospital for technical help during the writing.